# BotNet Intrusion Detection System in Internet of Things with Developed Deep Learning

Amirabas Kabiri Zamani, Amirahmad Chapnevis

Department of Computer & Information Technology Engineering, Amirkabir University of Technology, Tehran, Iran

**Abstract**
The rapid growth of technology has led to the creation of computing networks. The applications of the Internet of Things are becoming more and more visible with the expansion and development of sensors and the use of a series of equipment to connect to the Internet. Of course, the growth of any network will also provide some challenges. The main challenge of IoT like any other network is its security. In the field of security, there are issues such as attack detection, authentication, encryption and the so on. One of the most important attack is cyber-attacks that disrupt the network usage. One of the most important attacks on the IoT is BotNet attack. The most important challenges of this topic include very high computational complexity, lack of comparison with previous methods, lack of scalability, high execution time, lack of review of the proposed approach in terms of accuracy to detect and classify attacks and intrusions. Using intrusion detection systems for the IoT is an important step in identifying and detecting various attacks. Therefore, an algorithm that can solve these challenges has provided a near-optimal method. Using training-based models and algorithms such as Deep Dearning-Reinforcement Learning and XGBoost learning in combination (DRL-XGBoost) models can be an interesting approach to overcoming previous weaknesses. The data of this research is Bot-IoT-2018.

**Introduction**
The world of information technology and computers is expanding daily. This development has led to the creation of new systems with a specific type of communication. One of these communications is Machine-to-Machine (M2M). In this type of communication, as a solution to move from single-purpose devices that receive data in the form of commands from an application on the Internet that allows the device with solutions, to multi-objective and applications move towards cooperation together. M2M communication with network structures can benefit from global standardization efforts, which include a number of cases. Among these cases, the following points can be mentioned [1]:
- ✓ Establishing standards for compatibility with devices and applications without localization or with minimal localization by the appropriate device ecosystem to reduce the time required for deployment.
- ✓ Provide an ecosystem that allows programs to easily share their information and experiences.
- ✓ Provide an environment in which secure communication takes place and users' privacy and confidentiality are maintained.

In today's world, M2M solutions abound and their architecture has not changed much since the 1970s. The Franhouver Institute for Open Communication Systems defined the definition of M2M systems as communication terminals independent of human interactive communication with mainstream networks or other terminals in order to automate services. Admittedly, M2M communication network changed dramatically since the 1970s and expanded many capabilities (such as 3GPP M2M communication), but M2M architectural solutions have remained relatively stable. A device associated in the field with a software program on the main network for a specific purpose [2].

The most important issue in any network is security. Security analysis in computer networks is divided into several main categories including layers, identity and location of objects, authentication and permission, privacy, protocols and light weighted encryption, software vulnerabilities, mobile operating systems and so on. The connections between some of these principles in the analysis of security in the computer network environment are interdependent and only the field of Wireless Sensor Networks (WSNs) is a different topic because of a particular type of computer network. Security must be considered in different computer networks layers including sensor, network, platform and application which can be considered as a secure structure in computer networks. Also, according to the reference [3], security considered management as a control panel and synchronization for communications which can be in infrastructure networks, urban transmission network, data network, FTTx network, dimensional networks such as WAN and MAN, ports and sensor connections to the network [12-1].

System protection against vandalism or theft of hardware, software and data is provided as a definition of computer security and cyber security is defined as data protection in the network. According to the traditional definition of security, the topic of security is defined based on the three elements of data confidentiality, data integration and data availability. There are a number of security restrictions on computer networks that include:
- ✓ Hardware-based limitations: include energy and computational constraints, memory constraints, and equipment-resistant packaging.
- ✓ Software-based limitations: Includes limitations of embedded software and dynamic security patches when updating a piece of software.
- ✓ Network-based limitations: include mobility, scalability, multiplicity of devices, multiplicity of communication media, multi-protocol network, and dynamic network synchronization network.

There are four common classifications of attacks on computer networks: Reconnaissance attacks, access attacks, denial of service attacks, and data manipulation attacks. In Reconnaissance attacks, the attacker first examines the network systems and services provided or vulnerabilities of the computer network, and then search to detect any unauthorized entry and collection into the computer network. In most cases, these attacks act as a real access or Denial of Service (DoS) attack. Exploration attacks can be somewhat likened to a thief covering a neighborhood and infiltrating vulnerable homes (such as vacant homes, open doors or open windows). This paper presents a deep learning approach [6-9] based on XGBoost method to provide IoT intrusion detection to detect BotNet attacks.

**Literature Review**
In 1980, the concept of intrusion detection system began with [14, 13]. In these approaches, this issue began with the introduction of the idea that auditory attachments have vital information that may be useful in tracking abnormal behaviors and understanding user behaviors. In fact, it was the beginning of a host-based intrusion detection system. In 1986, a model was published that revealed the information needed to develop a business intrusion detection system [15]. MIDAS implemented an expert system using P-Best and LISP in 1988 [17, 16]. In the same year, Haystack was implemented, which tried to reduce audit attachments using statistics [18]. In 1989, it was implemented in [18] as a statistical anomaly detector that produced rules based on statistical analysis and then used those rules to detect anomalies and developed the network security monitoring and combined intrusion detection system [19]. In [19] an expert intrusion detection system called SRI is presented, which is associated with two approaches. A statutory system based on statistics and a statistical anomaly detector based on Sun's workstations that was able to examine data at both the user and network levels.

In 1991, the distributed intrusion detection system, which included an expert system developed by researchers at the University of California, San Diego, and an expert system, was implemented by the Los Alamos National Laboratory's Integrated Computing Network. The common denominator of attacks on these networks is DoS, DDoS, SQL Injection, MIM, Brute-Force, and the like. For this reason, in the cases examined, we do not constantly refer to the name of the attacks, because it is out of the question and only the type of method used in the system and in some cases a critical view of the method and in some cases the advantages of the method. Be. In 1998, the Lawrence Berkely National Laboratory introduced a legislative language called Bro to analyze packets from the libpcap data set. In 2001, tcpdump was used to create profile profiles for classifications in the analysis of audit data and the probe's intrusion detection system.

In [20], the approach of intrusion detection system based on probabilistic neural network in wireless sensor network has been considered, which has achieved significant results by using classification operations with the help of this neural network. In [21], the parallel design method uses a radial base function neural network to detect DOS and PROBING attacks and a multi-layered perceptron neural network to detect U2R and R2L attacks or a total of 41 features for each type of attack using multiple processors to identify and classify. These attacks have been used, the main drawback of which is the high volume of CPU usage, which is being processed by four processors at the same time and will be very expensive, but the advantage of that time is fast execution with penetration detection and classification.

An examination of the intrusion detection system using Hamming neural network to identify attacks on the TCP protocol shows that if the number of data examined is less than 1000, it is 100% correct, but when the number of data examined is high. When it reaches 5,000, the error detection percentage will be 88%. This dramatic error occurs with increasing data and is not the correct way to detect attacks [22].

Classification of normal patterns and attacks in the intrusion detection system using a three-layer neural network with a stop-validation-continue approach, which reduces the capacity of the neural network to work with data and increases training time due to the high volume of data has been examined. Once the parameters of the neural network have been determined by training, the classification of a single data record is done in a short time. Two layers of neural networks are used to classify interconnected records, and in the third layer the result of the classification appears. One of the problems with using this method to classify intrusions is the high execution time and computational complexity [23].

In [24] presented a fast and secure intrusion detection system in collaboration with multidisciplinary networks (cotton) as a network intrusion detection system and host-based intrusion detection system. In this cloud detection system, packets are received from the network, then analyzed, and then a report is sent to the cloud manager for further analysis. This manager is actually the owner of the same cloud, that is, the cloud user, and in the absence of a cloud, the system works automatically which is known as the main advantage of the job. The analysis method is classified and analyzed based on the combined model algorithm of the nearest K-neighbor and neural network. NSL-KDD data sets were used to train and test the data. After receiving a report from the cloud infiltration detection system, the cloud service provider generates an alert for the user to save the user's files in his absence to prevent malware and malware, and then detects and prevents possible attacks.

A 2015 study by Science Direct in 2015 [25] looked at risky methods as well as several methods of attacking Internet of Things, especially flooding and deprivation of services. Science Direct Magazine also examined a botnet attack on Internet service providers in 2016 while using the Internet of Things. This attack leads to the loss of the Internet for users and objects connected to the Internet. Conti, Mauro, et al., in 2017 [27] review the opportunities and challenges in the field of security in the Internet of Things. This approach, which is a case study, tries to solve it and provide new solutions by considering a series of attacks.

In the Pan, Meng-Shiuan, and Yang, Shu-Wei paper in 2017 [28] a distributed and lightweight multi-segmentation routing protocol was used. Initially, nodes require a lot of computation to decide how many parts to send. If the network has security holes, multi-section paths will be very long, and in some areas there may even be loops. The purpose of this study is to address existing challenges.

Another study by Jin, Yichao, and colleagues in 2016 used the RPL protocol for routing objects on the Internet [29]. The research approach proposed by this research is a content-based approach in which paths are determined by content. Due to the routing of data related to conventional relay nodes for processing, a higher rate of data accumulation is created, which is the first challenge that can be solved with the research approach of this research. Delayed reduction is also another issue in this study.

In the 2016 article by Kharkongor, Carynthia et al., [30] reducing energy consumption in heterogeneous devices connected to the Internet occurs in this study. An SDN controller is also used as the central administrator to provide a secure network environment.

Another study by Krishna, G. Gautham et al., [31] in 2016, used the RPL routing protocol to reduce energy consumption and extend the battery life of sensors in a wireless sensor network whose sensors are connected to the Internet. Has been considered. Also, a general analysis of routing methods in energy loss networks, including wireless sensor network, with the aim of reducing energy consumption in the Internet of Things and considering the support in it has been presented.

In an article by Badenhop, Christopher W. and colleagues in 2017 [32], the Z-Wave routing protocol is safely used in an Internet-based network environment. Almobaideen, Wesam, and colleagues in 2017 [33] have proposed a geographical routing method in IoT-based health monitoring systems to secure and detect existing attacks.

One of the best review articles ever published by 2019 by AP da Costa et al., [34] provided an overview of machine learning structures and algorithms in the field of intrusion detection systems for denial of services attacks, denial of distribution services and other attacks. Based on this research, it has been shown that attacks on intrusion detection systems on the Internet of Things with evolutionary algorithms and overcrowding intelligence have been presented and developed with higher speed and accuracy than other machine learning methods. One of the best methods available in the field of intrusion detection and attack detection systems on the Internet of Things, with deep learning approach by Abeshu Diro, Abebe, and Chilamkurti, Naveen in 2018 [35]. High accuracy is the most important achievement of this research.

In [36] proposed a hyper graph genetic algorithm to optimize parameters and select the best features in intrusion detection as well as the use of a support vector machine for these detections. In [37] combined fuzzy genetic systems and pairwise learning to improve the detection rate of attacks in an intrusion detection system. One of the advantages of this method is that fuzzy logic, due to its uncertainty in detecting intrusions which made it easier to produce it using membership functions, language variables, tags, rules, and fuzzy levels. Also, in order to detect attacks, considered interesting approach to design for learning to interpret and capture by considering a state space and searching with an evolutionary approach.

In [38], the Ant Colony Optimization (ACO) algorithm used to detect intrusion. This article provided a fuzzy detection system by combining fuzzy logic with entropy to select features and then create a real-time environment for classifying and detecting these intrusions and optimize it with ant colony algorithm. In [39] presented a hybrid classification system based on swarm intelligence methods in intrusion detection systems. The combined method is the use of support vector machine methods with the K-nearest neighbor along with the Particle Swarm Optimization (PSO) algorithm. Support vector machines and K-nearest neighbor used for classification based on distances and input data. Then particle swarm optimization algorithm used to create weights in order to optimizing classification environment with the best accuracy. KDDCUP99 used as input dataset. In another study presented in [40], the particle swarm optimization algorithm used to improve the existing intrusion detection system. This structure used linear programming and to improve the classification of the MCLP method. The data used was also KDDCUP99.

In [41] Feed-Forward Deep Neural Network (FFDNN) used to train attack data from KDD CUP and NSL KDD datasets. Also, feature extraction method used to identify the features, reduce the dimensions and select the best features for better classification structure as Wrapper Based Feature Extraction Unit (WFEU). 22 features of attacks have been identified that detection and classification rate in wireless networks including cloud computing network, IoT, WSN and VANET were 77.16% to 88.10%. Also in 42], the use of deep learning presented. In this research, an IoT-based intrusion detection system using KDD CUP 99 datasets presented which provide an immigration learning model and optimal feature extraction which were the important parts of this research.

The application of deep learning to intrusion detection in multi-cloud environments and wireless networks has been discussed in [43]. This approach accelerated the provision of a decision management system for intrusion detection and provided real-time processing for attack detection. The deep learning model is Denoising Autoencoder (DA) that creates blocks of the deep learning structure. The accuracy of this method on KDD CUP and NSL KDD datasets estimated up to 95%. Also in [44] offered many deep learning techniques that combined with Reinforcement Learning (RL) methods. These methods include Deep Q-Network (DQN), Double Deep Q-Network (DDQN), Policy Gradient (PG), and Actor-Critic (AC). The use of two datasets, NSL KDD and AWID considered in this research. It is worth noting that the deep learning model is presented with the conventional and well-known Q-Learning method of reinforcement learning and the results represented that the DDQN model reflects the best result with high accuracy in wireless networks.

Considering the IoT cloud problem as X in IoX presented an important issue in the field of secure data transfer [45]. IoT with the help of the cloud is becoming more and more prevalent in society, for example at home and at work. Hence the Internet with the help of cloud is also known as Objects (IoE). While in such settings, data can be easily shared and played (for example between a device like Amazon Echo and cloud or like Amazon AWS) there are potential security considerations that need to be considered. Therefore, a number of security solutions presented in this article. For example, Search Encryption (SE) extensively studied because of its ability to facilitate search on encrypted data. However, threat models in most available search encryption solutions rarely considered malicious data owners and cloud-based cloud servers at the same time, especially in dynamic applications. In the real world, there are differences between the two parties, as each party accuses the other of some wrongdoing. In addition, efficient full update operations (e.g., modifying, inserting, and deleting data) are not typically supported in cloud-based Internet deployment. Therefore, this article presented a fair and dynamic data sharing framework called FairDynDSF in multi-proprietary settings. Search results can be verified with FairDynDSF with fair, multi-word search, and dynamic updates. This research also proved that FairDynDSF is safe against keyword guessing attacks, and its performance has been demonstrated by evaluating its performance using different datasets.

In [46], the use of the IoT for X in IoX is located in the security of power systems. A true random number-based pseudohysteresis controller (PHC) proposed to prevent power side-channel attack (PSCA), power injection attack (PIA) and electromagnetic interference (EMI) for high-security devices on the IoT. In addition, the real-time quasi-hysteresis controller uses an advanced security constructor random value to generate a random number independent of the input power in the side channel power attack and power injection attack, leaving the random number to a residual window which converted the switch frequency constitutes a real random modulation.

Review articles in the field of IoX and IoT security are rarely provided which may be due to its prevalence [47]. In [48], an overview of network technologies studied on the IoX along with its applications for computing systems, and a look at security and intrusion detection. Also in [49], discussed and reviewed security challenges and IoX intrusion detection systems. The study of security and Block Chain convergence discussed as an important issue in [50] for the IoT. This security issue provided an overview of the challenges and future of Internet of Multimedia Things (IMoT) applications using Block Chain-level encryption systems and algorithms. In [51] sufficient security research has been done on individual elements on the IoX. The distinguishes article from the work of the IoTs and the IoX is that the IoT is treated as a component of an ecosystem in particular objects, and the threat model in a more comprehensive context than how other parts fit. The equation like individuals and data as well as processes developed. In [52] also discussed the issue of long-term sustainability for IoT security and the consideration of organic matter for communities. In general, the classification of intrusion detection systems is given in Figure (1).

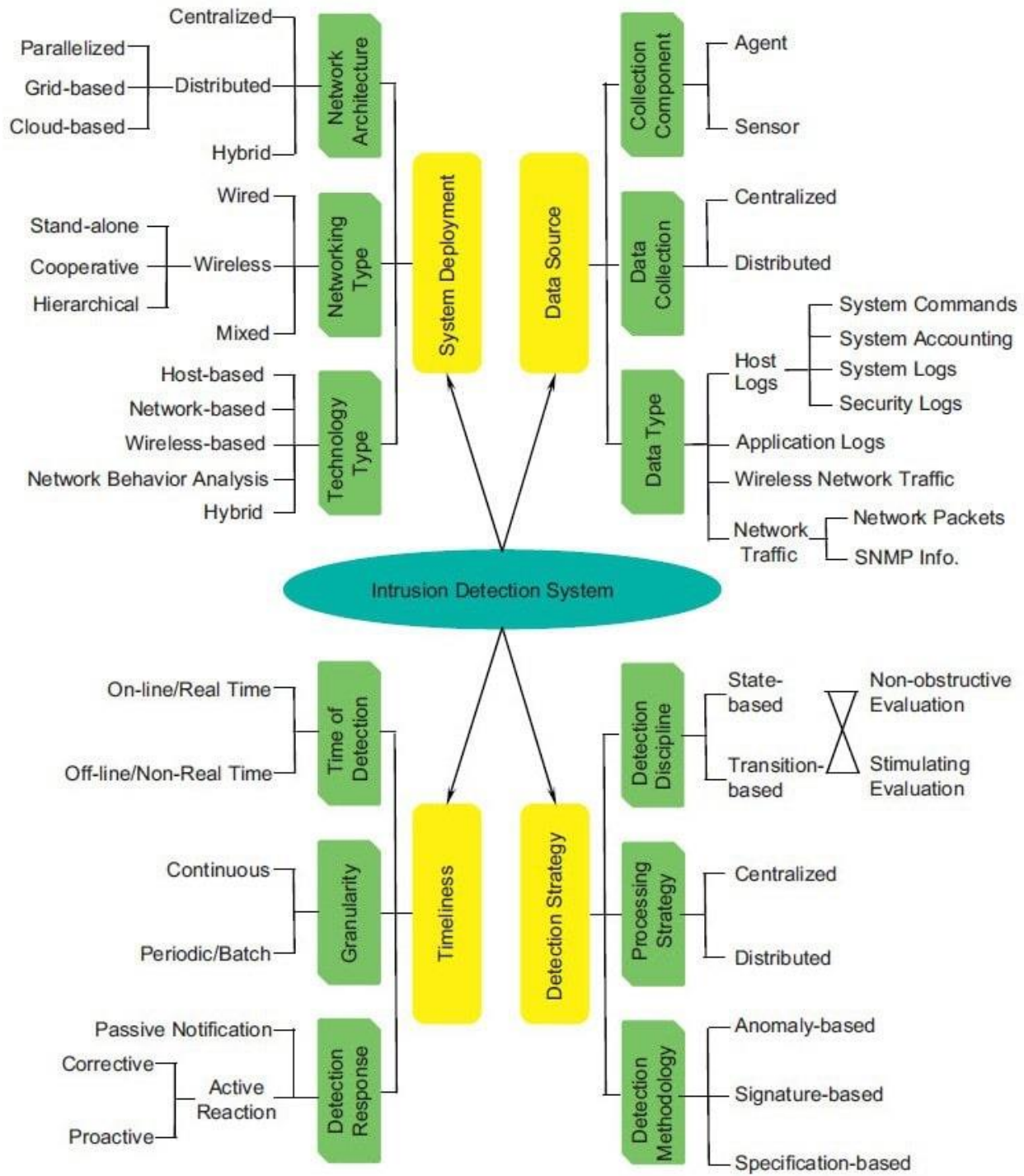

*Figure (1), classification of intrusion detection systems*

**Proposed Method**
The NSL-KDD data set is used to build a network-based intrusion detection system on the Internet. The research variables are in two categories: one is related to the types of attacks that can be detected, which is defined as Boolean, and using the Bot-IoT-2018 data set to detect intrusion into computer networks on the Internet that have attacks such as BotNet. The other category is about the efficiency of the intrusion detection algorithm, which can be expressed in terms of the duration of intrusion detection and the percentage of error in detecting intrusion and so on. MATLAB software is used to simulate the proposed approach, and the reason is that algorithms can be easily implemented in it. In order to ensure the proposed approach, several evaluation criteria will be used, including accuracy, sensitivity, mean error squares, bit error rate when preventing or detecting an attack, signal-to-noise ratio, and other methods used in other articles. In general, the proposed approach can be presented in the following phases:
- Select the appropriate data set
- Determine the type of attack and provide a model for their detection (include BotNet attack)
- Build a network environment in specific dimensions with the number of nodes
- Provide an intrusion detection system based on the proposed method, including the following steps and placing data in it.
- Normalize input data
- Data training and testing with the aim of classifying that will lead to the discovery of knowledge from the data.
- Select and extract the best features with XGBoost algorithm and data penetration detection
- Training data and testing with Deep Learning.
- Use evaluation criteria to ensure the proposed approach and compare with previous methods.

The main reasons for using the Deep Learning-Reinforcement Learning XGBoost (DRL-XGBoost) algorithm include the following:
- Has analysis with training on the process of attack and infiltration and creating rules to detect suspicious activities.
- Identify and identify additional consuming traffic that is suspicious.
- Prioritize alerts by identifying and marking alerts with lower priority.
- Identify scanned activities known at the network level and its activities.
- Analyze activities and provide reports to the team in response to network events and process monitoring by marking repetitive attack activities at the network level.
- Identify and detect unknown attacks using generalizability.

Networking in the Internet environment requires modeling. Initially, an environment is created in specific dimensions and a number of nodes are placed in it. Positioning refers to the initial deployment that is considered in two ways, either randomly or predefined. In this study, nodes are randomly located in the environment. An intrusion detection system is then provided with the XGBoost algorithm for classification, which has the ability to detect intrusions in a network environment on the Internet. Features will be based on the data in the data set and include the dimension reduction, selection and extraction of features. The classification also includes the placement of each attribute in its class (meaning that it is specified in the class or category of intrusions, without intrusion or suspicion).

Once the nodes are set up in a networked environment on the Internet, the data set that has a series of attacks needs to be given to the intrusion detection system on the Internet. Bot-IoT-2018 dataset is normally given to the intrusion detection system, and at the time of starting the network to send and receive data through nodes, all information exchanged in the network, with the intrusion detection system based on the available data from The NSL-KDD data set is examined and tested. But there is a need for a way to penetrate the intrusion detection system to improve its performance. To do this, the XGBoost algorithm begins to work on the feature extraction operation. When the volume of data is large, it is necessary to perform the property extraction operation, which has three main steps within itself, which include reducing the dimensions, selecting the features, and finally, extracting the feature (s). The features are then classified in the XGBoost algorithm.

The most important variables used are attacks on computer networks on the Internet, such as DoS, U2R, R2P, and DDoS. Identifying other variables of computer networks in the Internet to create its structure, including the number of nodes and servers, dimensions of the supported environment in the Internet environment, data distribution rate on the Internet and energy available throughout the network, are other important parameters.

The proposed method is based on the host-based intrusion detection system as well as the network-based intrusion detection system. The factor of host-based intrusion detection system is the factor of host-based intrusion detection system. This factor detects incompatibility in the cloud environment by deploying on the hosts by monitoring the behavior of the system files used in the hosts, network events, and system calls. It should be noted that in each host, a host-based intrusion detection factor is placed, which over time continues the learning process (described below) and maintains its efficiency.

The overall structure of this factor is very similar to the factors of the network-based intrusion detection system described below, the only major difference being that the agents monitor and locate the agent. Be. Explanatory notes on host-based intrusion detection system agents are described in detail, and the repetition of content for network-based intrusion detection system agents is avoided, only the differences are noted. The way in which a network-based intrusion detection agent works in network intrusion detection is, in terms of the detection process, exactly the same as that of a host-based intrusion detection system, except that instead of files. The system used in hosting, networking events, system calls to detect intrusion detection, controls packets.

But another major difference between the network-based intrusion detection system and the host-based intrusion detection system is the location of these agents in the cloud environment. In the proposed method, these agents monitor the traffic passing through each switch. This increases the computational overhead, but also slightly overlaps the functions between network-based intrusion detection system and host-based intrusion detection system, but in fact from another perspective and level. Attempts are being made to detect infiltration, as well as to provide additional coverage to the work of agents that increase the level of system reliability. Factor performance has two main phases:

Determining priority levels

Decide on new data

In the priority level determination phase, the determinant determines what data belongs to which priority level. In the decision-making phase, the threat of new data is identified and decisions are made about their level. The proposed method gives the system the flexibility to make different decisions at a priority level, although they are not very different from each other. Data classification is used to determine priority levels. In fact, the proposed method allows the valid methods of artificial intelligence, which is the use of XGBoost to be divided into different groups based on the actual behavior of the data.

The proposed system receives the data online according to the probability cycle. Then, by comparing these data with the centers of different classes, it determines the degree of importance or priority of this type of data. These priorities can vary from the lowest value that is a sign of non-threat data to the highest value that expresses the most threat. But as stated, we believe that flexibility can be achieved at any level of priority. This decision is communicated to the system and then recorded in the system history so that it can be used again in certain time periods to teach the system and its efficiency can be maintained over time.

It should be noted that XGBoost is used to make a more detailed decision about the data in each group. In this case, it is examined which of the possible rules in XGBoost are appropriate according to the specific conditions that will be described in detail below, and decisions are made according to them.

Once the data is idealized and the desired features are identified, XGBoost-based classification probabilities are performed. With the help of classifiers, data is classified into classes with the same characteristics. In general, classifiers are divided into two groups of supervisors with and without supervisors. In classification with the supervisor, the data are labeled and their affiliation is eloquent. In classification without observers, donors do not have a label. In this study, using the XGBoost classification method, data are classified into three categories, which include penetrating data, healthy data, and suspicious data penetration. Therefore, the method used in this project is classification with the supervisor. For this purpose, XGBoost classification is used to determine probabilities. The XGBoost classification relation is expressed as equation (1).

$$l_r(x) = \frac{p(x|\omega_1)}{p(x|\omega_2)} > \frac{p(\omega_2)}{p(\omega_1)} = x \in class \tag{1}$$

In this case, the function $p(x|\omega_1)$ indicates the conditional density of the corresponding class and $p(\omega_1)$ indicates the posterior probability of each class. Using this classifier, the pattern of lung changes is classified into three classes: infiltrated data, healthy data, and infiltrated suspicious data. The general algorithm works as follows:

- ✓ Step 1) The first step is to formulate the parameters that show the characteristics of the signals with $S$ and the risk criteria or risk factors that are with $F$, which are carefully adjusted based on an expert and have equation (2).

$$P = \{x | x \in R \cup S\} \tag{2}$$

- ✓ Step 2) Collect data that deals with whether the data is intact or not. Penetration data is shown as a set $D$ which is as equation (3).

$$\{d_1, d_2, \dots d_n\} \text{ where } d_i = \{p_{i1}, p_{i2}, \dots p_{im}\} \tag{3}$$

- ✓ Step 3) the data is thus in the pre-processing stage of classification: filling in the missing values between the data and the established states, converting continuous numerical variables to discrete variables using multi-digit thresholds. Inputs to definite inputs are filtering with a multiple filtering approach and data normalization.
- ✓ Step 4) the pre-processed data in the classification is placed in a set. For each confusion matrix model that contains the actual value, the actual value is negated and collected on the floor.
- ✓ Step 5) all the data characteristics and each classified output are placed in its own area, which consists of three areas, ie three floors, which include penetrating data, healthy data and suspicious data.
- ✓ Step 6) Develop the model based on the results found to reduce the dimensions.

After extracting features and classifying with XGBoost, it is necessary to perform XGBoost to combine the two sections, as well as to determine the probabilities of network intrusion detection. After determining the nearest class and focusing on high-similarity data to the new data, we then make a decision that we consider appropriate for the new data. The reason why the closest number of data is considered is that one can vote between them. Also, the reason why we choose more than one data as the closest data effective in decision making has been the reduction of the effects of noise or remote data. But how to tell if a data item is voting for a decision is not just about proposing a particular decision, and vice versa. Rather, the behavior is that if we make the decision $d$ for the $l_k$ data, a rule is generated in the form of XGBoost rules, and then we measure its strength by finding the probabilities. It is not possible to calculate the probabilities among the whole data set, but only the data in the same class is used to calculate these values. The probabilities for the XGBoost classification section are related to the rule $R = X \rightarrow Y$ such as equation (4).

$$P_{XGBoost}(R) = \frac{P(XY)}{|D|} \tag{4}$$

And the probabilities for the XGBoost classification section are in the form of equation (5).

$$P_{XGBoost}(R) = \frac{P(XY)}{P(X)} \tag{5}$$

In the above formulas, $P(.)$ is equal to the number of data from the whole set $D$ in which both $X$ and $Y$ are present. In our proposed method, $D$ is equal to the class from which the closest data was selected. But another important point that has not yet been mentioned about these rules is how to produce them and calculate the probabilities according to the type of data considered, which are discussed below.

It was previously stated that according to each of the $l_k$ data from the nearest set of properties, a rule is generated in the form of XGBoost. If the properties of $l_k$ are divided into two parts, decision and properties, then the whole set of properties can be displayed as $I_k = [f_i, \dots f_n, d_i, \dots d_m]$. In this representation, $f_i$ represents properties such as the type of connection and d_i represents the decision and accepts the value 0 or 1. Of course, only one decision has been made for each data, so only one of the $d_i$ to $d_m$ can be 1. For this data, all the rules are generated in the form of equation (6).

$R_1: f_1, \dots, f_{n_>} S_i$

$$R_2: f_1, \ldots, f_n \rightarrow S_i$$
.
.
.
$$R_{2n+1}: f_1, \ldots, f_n \rightarrow S_i$$
$$i \in [1, m]$$

(6)

As is clear from Equation (6), a decision may be made for several data. The evaluation of each law is done by calculating the probability values. In fact, the formulas given for calculating probabilities are appropriate for the data in the categories, and here the numerical data are difficult to make. The beauty of this powerful XGBoost algorithm lies in its scalability, which drives fast learning through parallel and distributed computing and offers efficient memory usage. XGBoost is an ensemble learning method. Sometimes, it may not be sufficient to rely upon the results of just one machine learning model. Ensemble learning offers a systematic solution to combine the predictive power of multiple learners. The resultant is a single model which gives the aggregated output from several models. The models that form the ensemble, also known as base learners, could be either from the same learning algorithm or different learning algorithms. Bagging and boosting are two widely used ensemble learners. Though these two techniques can be used with several statistical models, the most predominant usage has been with decision trees.

In boosting, the trees are built sequentially such that each subsequent tree aims to reduce the errors of the previous tree. Each tree learns from its predecessors and updates the residual errors. Hence, the tree that grows next in the sequence will learn from an updated version of the residuals. The base learners in boosting are weak learners in which the bias is high, and the predictive power is just a tad better than random guessing. Each of these weak learners contributes some vital information for prediction, enabling the boosting technique to produce a strong learner by effectively combining these weak learners. The final strong learner brings down both the bias and the variance.

In contrast to bagging techniques like Random Forest, in which trees are grown to their maximum extent, boosting makes use of trees with fewer splits. Such small trees, which are not very deep, are highly interpretable. Parameters like the number of trees or iterations, the rate at which the gradient boosting learns, and the depth of the tree, could be optimally selected through validation techniques like k-fold cross validation. Having a large number of trees might lead to overfitting. So, it is necessary to carefully choose the stopping criteria for boosting. The boosting ensemble technique consists of three simple steps:

- ✓ An initial model $F_0$ is defined to predict the target variable $y$. This model will be associated with a residual $(y - F_0)$.
- ✓ A new model $h_1$ is fit to the residuals from the previous step.
- ✓ Now, $F_0$ and $h_1$ are combined to give $F_1$, the boosted version of $F_0$. The mean squared error from $F_1$ will be lower than that from $F_0$.
- ✓ The mean squared error from $F_1$ will be lower than that from $F_0$ and calculated as equation (7).

$$F_1(x) \leq F_0(x) + h_1(x)$$

(7)

- ✓ To improve the performance of $F_1$, we could model after the residuals of $F_1$ and create a new model $F_2$ as equation (8).

$$F_2(x) \leq F_1(x) + h_2(x)$$

(8)

- ✓ This can be done for '$m$' iterations, until residuals have been minimized as much as possible like equation (9).

$$F_m(x) \leq F_{m-1}(x) + h_m(x)$$

(9)

Then output of XGBoost is input of deep learning algorithm. The type of deep learning is combination of deep learning with reinforcement learning (DRL-XGBoost. Different deep-reinforcement neural network models have been implemented, the only difference being the research approach in that two neural networks are used in DRL: one executes the current Q function while the other targets the Q function. The Q function is intended as a copy of the current Q function which is Q-Learning from the family of reinforcement learning algorithms, but works with a delayed coordination due to its presence and combination with a deep neural network. A copy is made after a certain number of training repetitions. The objective Q function is used to calculate the value of Q for the next case $(\hat{q}_t + 1)$. The purpose of this Q function is to prevent the effect of the moving target when performing a slope of more than $(\hat{q}_t - q_{ref})^2$ and to prevent the return of $q_{ref}$ dependence on the training network.

The algorithm begins by predicting actions using policy modes and functions. The action prediction is performed for all states of a path $(s_{\{t\}})$ and the sequence of predicted actions is generated. These predicted measures are obtained by sampling the probability distribution of measures $(\pi(a_{\{t\}}))$ provided by the policy performance. This section is considered as the probability of sample distribution. This research uses the symbol $\{T\}$ to indicate a sequence during the time steps of a path. When this symbol is used, it can have a sequence of scales, such as $r_{\{T\}}$ or a sequence of vectors such as $\pi(a_{\{T\}})$ or $\hat{a}_{\{T\}}$, because in the second case, $\pi(a)$ is the probability vector for any possible action under the current policy, and $\hat{a}_t$ is an encoded vector which assigned to the selected part, so extending them to a sequence produces vectors. The reward function creates a reward of 0/1, but in this case, it is a complete sequence of predicted actions $(\hat{a}_{\{T\}})$ and grand truth actions $(a^*_{\{T\}})$ which applied in one direction. The resulting bonus sequence $(r_{\{T\}})$ is converted to the vector of discounted bonus amounts $(R_{\{T\}})$. $(R_{\{T\}})$ is calculated by relation (10).

The policy gradient is based on a policy performance tutorial called $Q_{current}$ and using these two determines the operation that must be performed for each possible case. The policy function is performed with a simple neural network with multiple layers and ReLU activation for all layers except the last layer which has a SoftMax activation which is a possible distribution of actions or $(\pi(a))$. ReLU is a linear function that separately positively inputs, otherwise it will have zero. This has become the default activation function for many neural networks because the model they use is easier to teach and often performs better.

$$R_{\{T\}} = \left[\sum_{i=0}^{T} \lambda^i r_{t+i}, \ldots, \sum_{i=T}^{T} \lambda^i r_{t+i}\right] = [\sum_{i=0}^{T} \lambda^i r_{t+i}, \sum_{i=1}^{T} \lambda^i r_{t+i}, \ldots, \lambda^T r_{t+T}]$$

(10)

This means that each term $R_{\{T\}}$ corresponds to the decreasing amount of consecutive discount bonuses. The proposed method, due to the use of reinforcement learning model, uses the reward / punishment model that is the basis of these methods. From the vector of the discounted bonus amounts $(R_{\{T\}})$, the average of the discounted bonuses in different paths $(b_{\{T\}})$ is subdivided, and as a result, the superiority vectors of $(A_{\{T\}})$ is obtained. The vector $b_{\{T\}}$ is also called the baseline. The vector means the amount of rewards and punishments that will be counted as a set of data. Advantage values estimate how much better the expected return for a particular element of the path $(s_t)$ is than the average expected return. This is the reason for subtracting the baseline from $R_{\{T\}}$.

The scalar product between the sequences of the vectors $\pi(a_{\{T\}})$ and $\hat{a}_{\{T\}}$', derives the probability of a selective action for each time step $(\pi(\hat{a}_{\{T\}}))$, because $\hat{a}_t$ is an encrypted vector. The loss used to train the neural network which is an approximation of the policy function is a type of log-loss function with the sum of the recorded paths of the probability of action performed for a particular element of the path $(\log \pi([\hat{a}_{\{T\}}]_i)$ multiplied by the value of the corresponding advantage $([A_{\{T\}}]_i)$. When the training is completed, the neural network that executes the policy function is used to predict. The specific case provides the functionality of the probability distribution policy for actions. Possibility distribution is an important part of reinforcement learning that in a combined model with deep learning which can determine the type of probability to achieve the result and its distribution, structural superiority. In this prediction mode, only the operation with the highest probability (without sampling) is selected. The output of each previous layer and its bias with the nonlinear actuator function $f$ to generate weighted input $W_n$ for the next layer $n$ of a neural network is calculated in Equation (11), the loss function is the first part in identifying errors and intrusions using the mean squared error for one $k$ of training data, and the second part is to avoid overfitting during training. Overfitting is a model that formulate training data. Overfitting occurs when a model trains details and noise in training data to the extent that it negatively affects the performance of the model on new data. This means that random noise in training data are selected and trained by the model as concepts. The problem is that these concepts do not apply to new data and negatively affect the modeling ability to generalize. Overfitting is more in non-parametric and nonlinear models that have more flexibility when learning target performance. Similarly, many non-parametric machine learning algorithms also include parameters or techniques for limiting the amount of model details.

$$J(W,b) = \frac{1}{2\square} \sum_{k=0}^{k} (||x^{(i)} - \hat{x}^{(i)}||^2 + \frac{\lambda}{2} \sum_{l=1}^{nl-1} \sum_{i=1}^{sl} \sum_{j=1}^{sl+1} \left(W_{ij}^{(l)}\right)^2$$

(11)

In this regard, $nl$ represents the number of layers in the deep neural network and $sl$ is the number of neurons in each input layer. By combining equation (10) and (11), a structure is presented as a combination of deep-reinforced neural network, the general equation is as (12).

$$R_{\{T\}}.J(W,b) = (||x^{(i)} - \hat{x}^{(i)}||^2 \cdot \sum_{i=1}^{T} \lambda^i r_{t+i} + \frac{\lambda}{2} \cdot \left(W_{ij}^{(l)}\right)^2$$

(12)

An accurate diagnosis of any intrusion and suspicious symptoms can be provided with the help of this equations during training and testing. Also, the structure of the IoT network is an $N_x \times M_y$ environment which will be in terms of square meters in which the number of nodes or $N_{user}$ in different environments are randomly located by devices connected to the Internet. Any transmitting and receiving of data can be controlled once the routing operation is started and monitored with the proposed approach as an intrusion detection system based on the proposed DRL algorithm. To ensure the discussion of energy when detecting intrusions, the issue of heat in the processor is raised which is in the base station. The energy consumption during the cycle can be estimated based on the amount of energy consumption of the nodes to receive or transmit data at the same time as identifying and detecting intrusions in each cycle. A first-class radio model is used to measure energy consumption. The energy used to transmit a one-bit packet from transmitter to receiver at a distance $d$ at the same time as intrusion detection can be defined as Equation (13).

$$E_{TX} = \begin{cases} lE_{elec} + l\varepsilon_{fs}d^2, & d < d. \\ lE_{elec} + l\varepsilon_{mp}d^4, & d \geq d. \end{cases}$$

(13)

In this regard, $E_{elec}$ is the scattered energy to work with the transmitter or receiver circuit per bit, $d$ is the transmission distance. $\varepsilon_{fs}$ and $\varepsilon_{mp}$ are the amplifier energy factors for open space and the multi-path dimming channel models, respectively. The intersection $d.$ is the threshold distance that depends on the specific scene and the amplifying energy factors, which can be given as $d. = \sqrt{\varepsilon_{mp}/\varepsilon_{fs}}$. The energy used to receive one-bit data can be written as equation (14).

$$E_{Rx}(l,d) = lE_{elec}$$

(14)

And the energy consumed for the aggregated data is also in the form of equation (15).

$$E_{Agg}(l,d) = lE_{DA}$$

(15)

In this regard, $E_{DA}$ is the energy used to send the accumulated data bit. It is necessary to balance the energy between the energies of the sensor nodes to extend the life of the network at the time of intrusion detection. Here, the additive learners do not disturb the functions created in the previous steps. Instead, they impart information of their own to bring down the errors. DRL-XGBoost is a popular implementation of gradient boosting. Let's discuss some features of DRL-XGBoost that make it so interesting.

- ✓ Regularization: DRL-XGBoost has an option to penalize complex models through both L1 and L2 regularization. Regularization helps in preventing overfitting.
- ✓ Handling sparse data: Missing values or data processing steps like one-hot encoding make data sparse. DRL-XGBoost incorporates a sparsity-aware split finding algorithm to handle different types of sparsity patterns in the data.

Weighted quantile sketch: Most existing tree based algorithms can find the split points when the data points are of equal weights (using quantile sketch algorithm). However, they are not equipped to handle weighted data. XGBoost has a distributed weighted quantile sketch algorithm to effectively handle weighted data.

Block structure for parallel learning: For faster computing, DRL-XGBoost can make use of multiple cores on the CPU. This is possible because of a block structure in its system design. Data is sorted and stored in in-memory units called blocks. Unlike other algorithms, this enables the data layout to be reused by subsequent iterations, instead of computing it again. This feature also serves useful for steps like split finding and column sub-sampling.

Cache awareness: In DRL-XGBoost, non-continuous memory access is required to get the gradient statistics by row index. Hence, DRL-XGBoost has been designed to make optimal use of hardware. This is done by allocating internal buffers in each thread, where the gradient statistics can be stored.

Out-of-core computing: This feature optimizes the available disk space and maximizes its usage when handling huge datasets that do not fit into memory.

Several evaluation criteria have been used in this study, including Mean Square Error (MSE), Peak Signal-to-Noise Ratio (PSNR), Signal-to-Noise Ratio (SNR), and precision criteria. At the beginning, each survey is completed and finally the results of the research performed with each evaluation criterion are mentioned. The accuracy rate is a criterion expressed as a percentage, which is the most important overall result of the evaluation criteria section, which is the accuracy of the relationship and its equation is (16).

$$Accuracy = 100 \times \frac{TP + TN}{TN + TP + FN + FP}$$

(16)

In equation (16), $TP$ is false positive, $TN$ positive negative, false positive $FP$ and false negative $FN$. Equation (17) shows the sensitivity expressed in percentage.

$$Sensitivity = \frac{TP}{TP + FN}$$

(17)

Equation (18) shows the data properties expressed in percentage.

$$Specificity = \frac{TN}{TN + FP}$$

(18)

**Simulation and Results**

The simulator used in this research is MATLAB. One of the reasons for its use is due to the simplicity of using smart methods and algorithms that have already been coded and need to be coded and modeled according to the problem. But there are other simulators that can be used. Among these simulators, the following can be mentioned that the problem of using each of them is expressed separately (although each of them also has advantages that can be ignored):
- ✓ CloudSim: The ability to use evolutionary methods and swarm intelligence is difficult and must be coded in C ++ and called as a library in this simulator. It also requires a series of functions in the form of header files with the extension .h.
- ✓ NS-2 or NS-3: Installing these two emulators is very complicated and difficult, and with the functionality of an intrusion detection system, it requires the installation of a series of separate packages as plugins. The code must also be written in tcl and the main parts of the proposed algorithm must be written and added in C ++.
- ✓ OPNet: This simulator does not have any interesting free versions in Iran and using it is a high risk of endangering the proposed methods as an idea in a variety of networks.

- OMNet ++: A powerful emulator, but its installation is complex and requires expressions to launch, and the code of the proposed method must be written and added in C ++.
- Other emulators, including JSIM, QualNET, and GNS3, lack the libraries needed to simulate this approach, using cloud computing networks and intrusion detection systems, and all sections must be coded and integrated in C ++ and Java, which is complicated. And it takes time.

Bot-IoT-2018`as dataset used in this approach. First, it is necessary to define the basic parameters of the computer network on the Internet. Table (1) shows the parameters of a computer network on the IoT.

*Table (1) Computer network parameters in the IoT*

| Network Scale | 100x100 m2 |
|---|---|
| Nodes or Users Numbers | 200 |
| Nodes or Users Distribution Rate | 0.05 |
| Nodes or Users Energy | 0.5 |
| ransmission Energy | 20 Joule |
| Radio Range | 60 m |

The data used in this research is Bot-IoT-2018 which has different versions. The version used in this research is the 2018 version, which has attacks such as DoS, U2R, R2P, SP and AUB, but we will use BotNet attack to detect and classify. The placement of nodes, which are also moving is done randomly in the environment, which is neighborhood and distance based on Euclidean distance, based on radio radius. This can be seen in Figure (2).

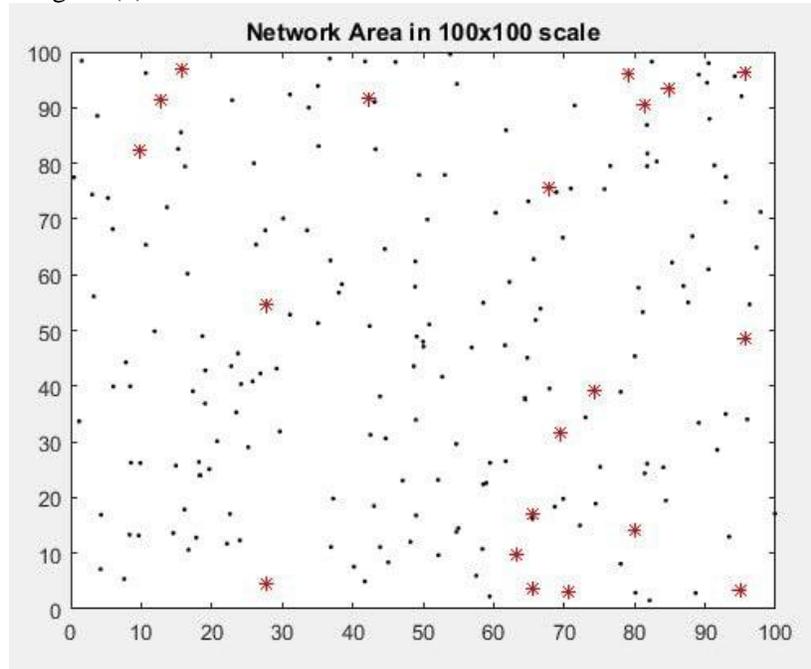

*Figure (2), network dimensions and node deployment as well as DRL-XGBoost decision priority selection for packet distribution*

Then, in Figure (3), the network life is shown as a signal, which is based on different network repetitions in penetration detection. Circles indicate the peak or highest and lowest energy consumption.

---
`https://research.unsw.edu.au/projects/bot-iot-dataset

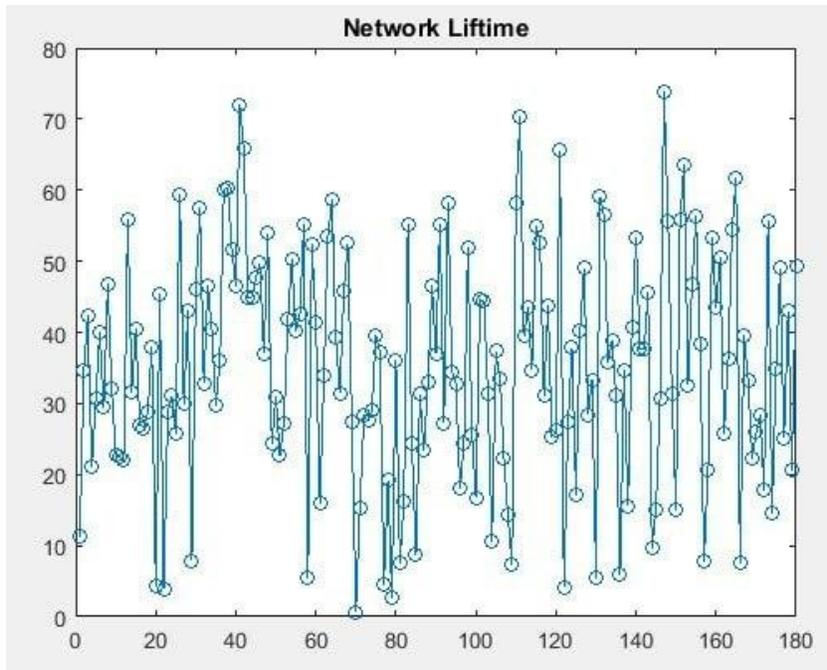
*Figure (3), network life as a signal with the highest value and the lowest energy consumption*

In Figure (4), the proposed method can be applied to detect penetration and prevent it. The red graph shows this. In areas where there is a square (blue) and green circles inside it, it shows the prevention of intrusion in that area in terms of signal-to-noise ratio, which is based on network life and stability.

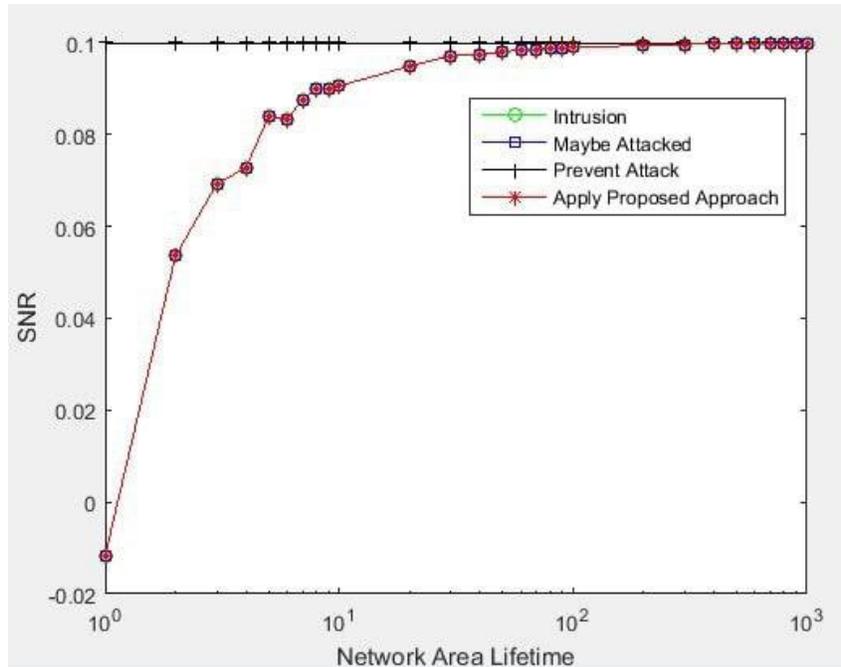
*Figure (4), apply the proposed method to detect penetration and prevent it*

Figure (5) also shows the reduction of intrusions and attacks in the Internet-based computer network environment, which are detected and identified by the approach presented in Chapter Three. The red graph shows this. In areas where there is a square (blue) and inside the green circles, it shows the prevention of penetration in that area in terms of longevity and stability.

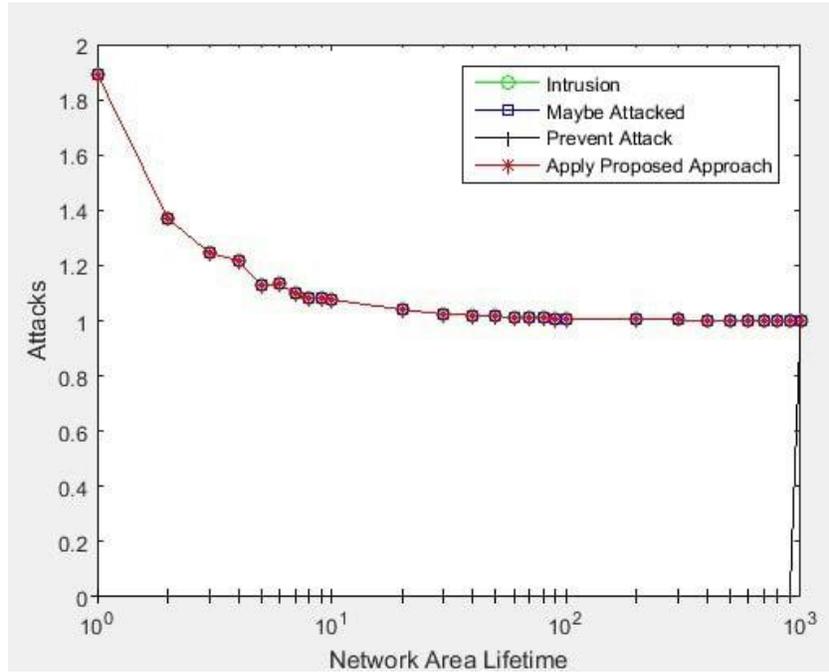

*Figure (5), reduction of intrusion and attacks in the network after applying the proposed method*

After implementation, the accuracy of the proposed method in detecting and preventing intrusion is 99.9940%. Also, the average error square is 0.1315. Table (2) compares the proposed interstitial accuracy with four other studies.

*Table (2) Comparison in terms of accuracy between the proposed method and two other methods*

| Methods | Accuracy |
| --- | --- |
| [43] | 96.032 % |
| [20] | 89.025 % |
| [21] | 88.700 % |
| [44] | 97.59 % |
| Proposed Method | 99.994 % |

According to the results, it is clear that the proposed method has a functional advantage over the other two similar and previous methods in terms of accuracy in intrusion detection.

**Conclusion**

Security is an important issue for the IoT as it is for other networks. In this research, IoT components are initially initialized including sensor nodes, network dimensions, and intelligent physical objects connected to the Internet. The sensor nodes are the users that will transfer the data. Data transfer will be called routing to send packets from one source to another destination. The data packet is placed at the source, at the time of transmitting on the communication channel and the IoT platform where the intrusion detection system targets data packet with trained data. Bot-IoT-2018 data used as input dataset in this research and the main attack is BotNet. This data has been trained once in the DRL-XGBoost intrusion detection system. Therefore, after transmitting the data by the source, it is placed in this area as a monitoring area. The pattern of possible attacks in user-submitted data is compared in pairs with the patterns trained in the DRL-XGBoost based intrusion detection system, and in case of any intrusion, the attack detection module is activated and the operation is performed that can temporarily stop transmitting data from source to destination. In fact, such a system ends when the operations delivers the message to the IoT. Therefore, this system does not have a meaningful termination, i.e. it may be sent in pseudo-continuous data and should be constantly checked, but logically, the termination part will be after the intrusion detection system.

[45] Yinbin Miao, Ximeng Liu, Kim-Kwang Raymond Choo, Robert H. Deng, Hongjun Wu, and Hongwei Li. (2019). Fair and Dynamic Data Sharing Framework in Cloud-Assisted Internet of Everything. IEEE Internet of Things Journal, Vol. 6, Issue: 4, pp. 7201-7212.

[46] Wen-Hau Yang, Shao-Wei Chiu, Chun-Chieh Kuo, Yen-Ting Lin, Yan-Jiun Lai, Hung-Wei Chen, Yu-Sheng Ma, Ke-Horng Chen, Ying-Hsi Lin, Shian-Ru Lin, and Tsung-Yen Tsai. (2020). A True-Random-Number-Based Pseudohysteresis Controller for Buck DC–DC Converter in High-Security Internet-of-Everything Devices. IEEE Transactions on Power Electronics, Vol. 35, Issue: 3, pp. 2969-2978.

[47] Souvik Pal, Vicente García Díaz, and Dac-Nhuong Le. IoT: Security and Privacy Paradigm (Internet of Everything IoE). CRC Press; 1st edition (25 Jun. 2020), 399 pages.

[48] Mohammad Shojafar, and Mehdi Sookhak. (2019). Internet of everything, networks, applications, and computing systems (IoENACS). International Journal of Computers and Applications, Vol. 42, Issue 3, pp. 213-215.

[49] Melnik Sergey, Smirnov Nikolay, and Erokhin Sergey. (2017). Cyber security concept for Internet of Everything (IoE). IEEE 2017 Systems of Signal Synchronization, Generating and Processing in Telecommunications (SINKHROINFO), Kazan, Russia.

[50] Mian Ahmad Jan, Jinjin Cai, Xiang-Chuan Gao, Fazlullah Khan, Spyridon Mastorakis, Muhammad Usman, Mamoun Alazab, and Paul Watters. (2020). Security and blockchain convergence with Internet of Multimedia Things: Current trends, research challenges and future directions. Journal of Network and Computer ApplicationsAvailable online 27 November 2020, 102918, In Press, Journal Pre-proof.

[51] Jungwoo Ryoo, Soyoung Kim, Junsung Cho, Hyoungshick Kim, Simon Tjoa, and Christopher Derobertis. (2017). IoE Security Threats and You. IEEE 2017 International Conference on Software Security and Assurance (ICSSA), Altoona, PA, USA.

[52] Kazunori Kuribara, Taiki Nobeshima, Atsushi Takei, Takehito Kozasa, Sei Uemura, and Manabu Yoshida. (2019). Long-term stability of organic physically unclonable function for IoE security. 2019 Compound Semiconductor Week (CSW), Nara, Japan, Japan.
19